\documentclass[journal=apchd5,email=true]{achemso}

\usepackage{graphicx,color,framed,helvet}

\title{Total Internal Reflection for Effectively Transparent Solar Cell Contacts}
\author{Phillip Jahelka}
\author{Rebecca Saive}
\author{Harry Atwater}
\email{haa@caltech.edu}
\affiliation{Division of Engineering and Applied Sciences, California Institute of Technology, Pasadena, CA 91125}
\begin{document}







\begin{abstract}
A new strategy for eliminating photocurrent losses due to the metal contacts on the front of a solar cell was proposed, simulated, and tested. By placing triangular cross-section lines of low refractive index on top of the contacts, total-internal reflection at the interface of the low-index triangles and the surrounding material can direct light away from the metal and into the photoactive absorber. Simulations indicated that losses can be eliminated for any incident angle, and that yearly energy production improvements commensurate with the metallized area are possible. Proof of principle experiments were carried out to eliminate the reflective losses of a commercial solar cell's busbar contact. Spatially resolved laser beam induced current measurements demonstrated that reflection losses due to the busbar were reduced by voids with triangular cross-section.
\end{abstract}


\maketitle

	In order to mitigate the effects of global warming, new energy sources that do not release greenhouse gases must be deployed to displace carbon intensive energy sources such as coal and oil.\cite{Edenhofer2014} Solar energy is a promising energy source due to the abundant energy received by the earth from the sun. In order to make solar economically competitive with fossil fuels, its cost must be reduced through some combination of increasing efficiency, which reduces the number of modules that must be installed, or by directly decreasing the cost to install a module. Increasing efficiency is an intriguing route to decrease cost because most commercial silicon solar modules still operate at less than 18\% efficiency; well below the detailed-balance efficiency limit of 29.8\% for silicon.\cite{Tiedje1984,Barbose2015}
	
	One fundamental problem limiting the efficiency of most silicon solar cells is reflection losses due to the metal contacts on the front of the cell.\cite{Luque2011} The contacts are a necessary evil because they are required to achieve low electrical resistance, but reflect light away from the solar cell, thus reducing the absorbed photocurrent. A number of strategies have been developed to minimize the reflection losses. These strategies can broadly be classified by whether they fundamentally modify the cell design, change the front contacting scheme, or rely on light management outside the cell. One established cell design that eliminates front metal reflection losses are interdigitated back contact cells.\cite{Lammert1977,Smith2012,kaneka} While interdigitated back contact cells completely eliminate the front metallization reflection losses and hold the current silicon efficiency records, they suffer from requiring a much more complex fabrication process. Strategies modifying the front metallization include the use of transparent conductive oxides (TCOs), fractal contacts, and high aspect ratio triangular contacts.\cite{Saive2016,Afshinmanesh2014,Koida2009,Yu2012,Kobayashi2016} TCOs seek to solve the problem by eliminating the metal, but suffer from high cost, intrinsic parasitic absorption, and low sheet conductivities. High aspect ratio silver triangular contacts work by directly reflecting incident light into the solar cell and have demonstrated over 99\% reduction in reflection losses and excellent sheet conductivities, but require additional metallization and have not been demonstrated on textured silicon solar cells. Finally, an existing strategy that relies on external optics is prismatic encapsulation where the front surface of the module acts as a lens directing light away from the contacts.\cite{Boca2009} Unfortunately, this method reduces the acceptance angle of solar radiation by actively concentrating the sunlight. 
	
	In this paper, we demonstrate a novel technique for eliminating solar cell front contact reflection losses by modifying the encapsulation design. Every solar cell must be encapsulated in rugged material to survive outdoors. By inserting regions with refractive index lower than the surrounding encapsulation and with suitable geometry above the contacts, incident light will undergo total internal reflection due to the contrast in refractive index and be reflected into the absorbing material instead of out of the module. Benefits of the proposed design include compatibility with commercial solar cells, independence of angle of incidence, and simple implementation. We first present a toy theoretical model to predict the behavior of such structures, followed by simulation results confirming the increase in photocurrent. Finally, we include proof of principle experimental results demonstrating the efficacy of this strategy.

\section{\label{theory}Theory}
Figure 1a illustrates the front contact reflection problem. The solar module is modeled as a superstrate of air, a thin sheet of glass, a layer of encapsulant with index of refraction $n_e$, a thin metal contact with perfect reflectivity, and a solar cell substrate with perfect absorption. The glass and polymer are assumed to have the same index of refraction. A light ray incident at angle $\theta_i$ strikes the module, is refracted by the air/glass interface, reflected by the metal, and then leaves the module. The escape means the cell is losing photocurrent to reflection off the metal contact. 
\begin{figure}[h!]
	\centering
	\def\svgwidth{\columnwidth}
	\begin{framed}
	
		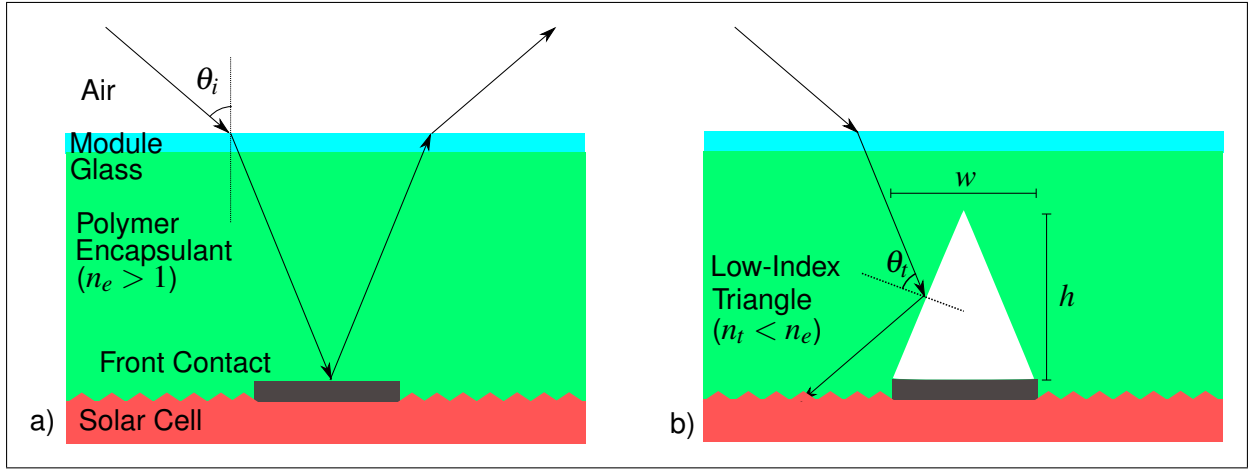
		\end{framed}
		\caption{Illustration of how low-refractive index regions can improve photocurrent of solar cells. (a) Demonstration of why metal contacts reduce photocurrent. Light incident on the solar cell reflects off the cell's front metal contact, reducing the photocurrent of the cell. (b) Example of how lower refractive index regions can improve photocurrent. A low refractive index triangle is placed above the metal contact. Assuming a favorable relationship between the angle of incidence, triangle geometry, and index contrast, the light reflects off the encapsulant/triangle interface and into the solar cell.}
		
\end{figure}

Figure 1b is our proposed solution to the reflection problem. By placing a triangle with index of refraction $n_t<n_e$, height $h$, and width $w$, and sufficient aspect ratio, the ray will reflect off the triangle/encapsulant interface due to total internal reflection (TIR). This reflected ray is then absorbed by the solar cell instead of escaping the module, thus increasing the absorbed photocurrent. Using Snell's law, simple constraints can be derived for when TIR will occur. For a given triangle geometry, indices of refraction, and incident angle, the angle of incidence of the light on the triangle/encapsulant interface, $\theta_t$ is:
\begin{equation}
\theta_{t}=tan^{-1}\left(\frac{2h}{w}\right)-sin^{-1} \left(\frac{sin\left(\theta_i\right)}{n_e}\right).
\end{equation}
Setting $\theta_{t}$ to the critical angle for the triangle/encapsulant interface and solving for $\theta_i$ returns the maximum angle of incidence for which TIR is supported:
\begin{equation}
\theta_{i,\mathrm{max}}=sin^{-1} \left[n_e sin\left(tan^{-1}\left[\frac{2h}{w}\right]-sin^{-1}\left[\frac{n_t}{n_e}\right]\right)\right].
\end{equation}
An interesting feature of this equation is that for $n_e/n_t>\sqrt{2}$, it becomes possible to choose $2h/w$ large enough that $\theta_{i,\mathrm{max}}=\pi/2$. This means that the light will be directed away from the contact for any angle of incidence. The physical interpretation of this result is that it is the index contrast at which a vertical boundary will reflect light traveling along the air/encapsulant interface's escape cone. Triangles that satisfy the constraints for TIR at $\theta_i=\pi/2$ are subsequently referred to as `perfect triangles.'

\section{\label{Sim}Simulation}

	To study the efficacy of the triangles, we performed a number of ray tracing simulations using LightTools. We performed two types of simulations. First, we modeled a single contact and studied how the absorbed photocurrent changed with angle of incidence and geometry of the structure above the contact. Secondly, we modeled a full solar cell and calculated the yearly energy harvest for various tracking strategies. In all simulations, the solar cell was modeled as a perfect isotropic absorber, the contacts as specular reflectors, the triangles were assumed to be air, and the encapsulation has index of refraction 1.55 to model standard solar cell encapsulants. 
	
	In the first type of simulation, the solar cell is 20mm on each side with ecanpsulant 6mm
	thick and 40mm per side on top. The contact is a 2mm wide strip running down the center.
	The ray source is a large sheet just above the cell whose angle of emission is controllable.
	We first simulated the photocurrent of a cell without a triangle and then simulated triangles with various geometries and normalized the photocurrents to the response of the cell without triangles.
	
	Figure 2a illustrates the setup and our convention for angle of incidence. A line parallel to the z axis is first rotated in the y-z plane by $\psi$ and then rotated in the x-z plane by $\omega$. The line is then parallel to the direction of incidence. The intent is that $\psi$ is the angle of incidence along the contact and $\omega$ is the angle of incidence perpendicular to the contact. Figure 2b contains the results of the simulations. In each subplot, the horizontal axis is the angle of incidence along the contact, $\psi$, and the vertical axis is the angle of incidence perpendicular to the contact, $\omega$. The color represents the relative photocurrent change compared to the plain cell. The geometry of the triangle is defined in the bottom left. The large red region is where the relative photocurrent improvement is 10\%. The 10\% improvement indicates the triangle is completely eliminating the reflection losses because the metal covers 10\% of the cell. However, moving to large $\omega$, performance suddenly degrades due to the failure of TIR; even performing worse than the plain cell. This is due to the failure of TIR inducing scattering of light within the module worse than the metal alone would cause. Moving one subfigure to the right we see that increasing the width of the triangle while keeping the height constant does indeed degrade the range of angles of incidence for which the triangle is effective. However, while TIR is working, we observe the same improvement in photocurrent. By increasing the height in the top-right figure, the angular performance is recovered. The top left plot is for a triangle with the minimum height to qualify as ‘perfect.’   The simulation validates that a perfect triangle will enhance the photocurrent for all angles of incidence. The fluctuations at extreme angles of incidence are due to numerical instability caused by the edge of the encapsulation becoming important.
	
	\begin{figure}[h]
		\centering
		\def\svgwidth{0.9\columnwidth}
		\begin{framed}
		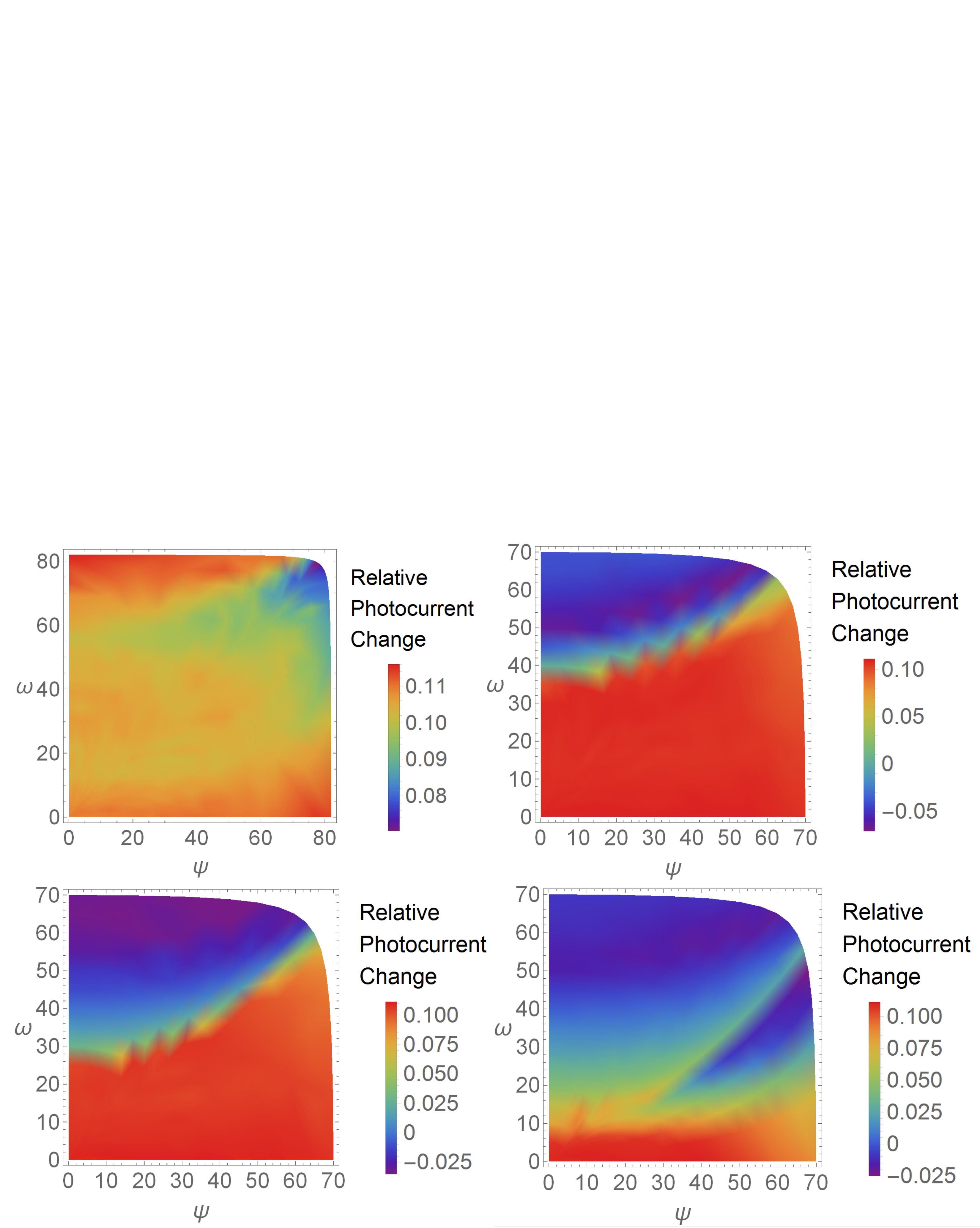
		\end{framed}
		\caption{Exploration of angle of incidence dependence of photocurrent change as a function of triangle geometry. (a) Schematic illustrating the simulated geometry and the scheme for encoding the angle of incidence. (b) The major horizontal axis represents increasing triangle width and the major vertical axis increasing triangle height. In each subplot, a point in the plane is the projection of the angle of incidence on the module. The color denotes the change in photocurrent compared to a plain cell. The geometry of each triangle is indicated in the lower left.}
	
	\end{figure}
	
	We also studied the properties of other shapes. Two we focused on were a sawtooth and ‘house.’ Figure 3a depicts the geometries and some sample light rays for the sawtooth and house. The sawtooth is interesting because the response depends on which side of the contact the light is incident from. When light is incident on the side with the vertical wall, TIR will be present for a very large range of incident angles, while light incident on the shallowly sloped side will experience TIR for a much smaller range of incident angles. The house is intriguing because the walls will provide TIR over a wide range of incident angles while the roof will support it over a narrower range. Figure 3b shows the simulations results for these structures. The top plot demonstrates the asymmetry of the sawtooth behavior where photocurrent is enhanced when light is incident on the vertical side, but degraded when incident on the shallow-sloped side. The house has very interesting behavior because when TIR fails on the roof, the overall photocurrent response is not degraded compared to the reference whereas it is degraded when TIR fails for a triangle. These is also a smoother transition between the regions where TIR works and fails. These shapes could be useful if the light source will be known to not be equally incident on both sides of the contact or if a weaker angle dependence is required and perfect triangles are not possible.
	
		\begin{figure}[h!]
			\centering
			\def\svgwidth{\columnwidth}
			\begin{framed}
			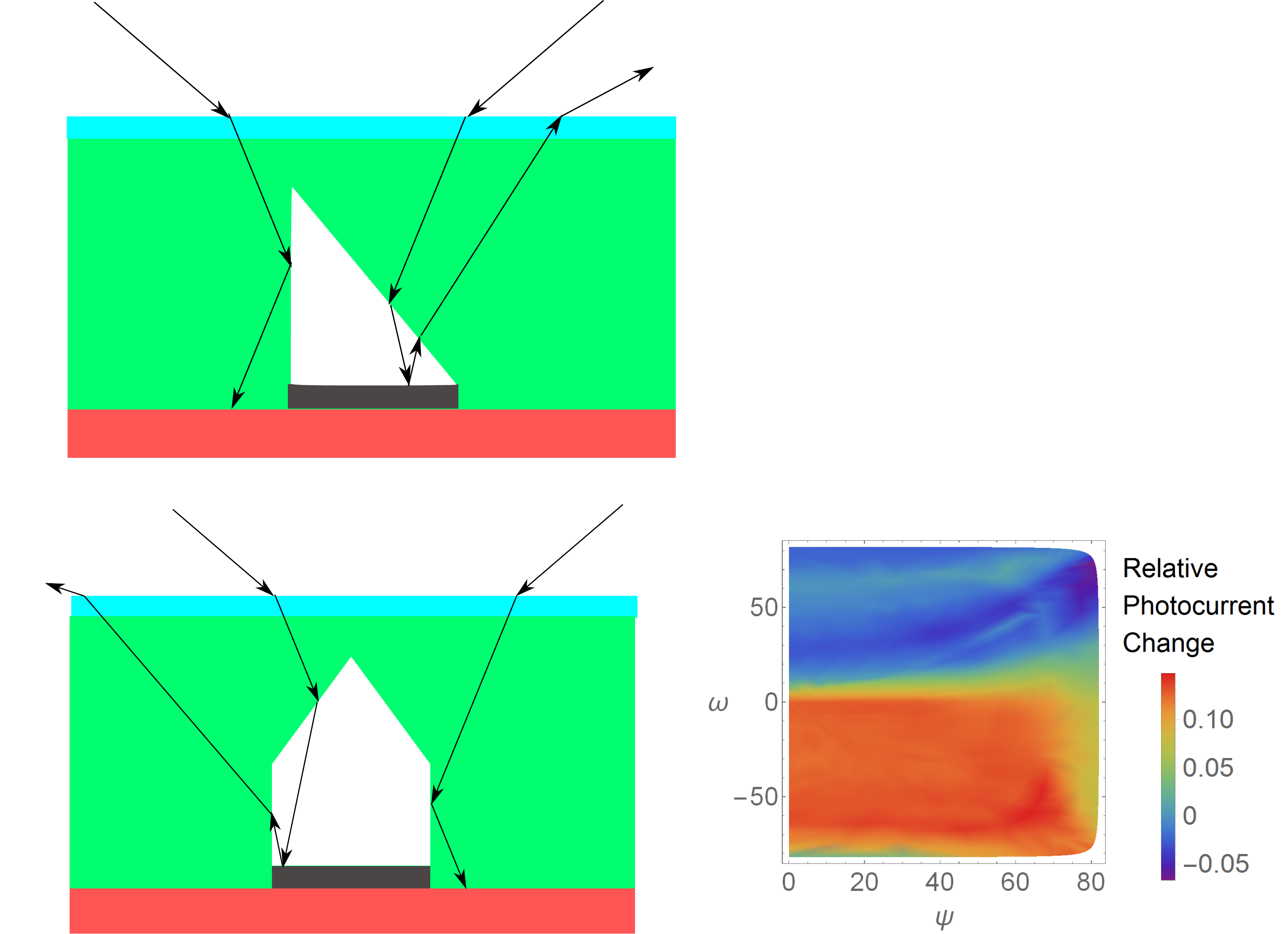
			\end{framed}
			\caption{Examining different structures and their performance as a function of angle of incidence. (a) Two possible non-triangular structures for the low-index region above the contact. The upper structure is a sawtooth, and the other is a `house.' The walls and roof of the house are each $2\mathrm{mm}$ tall. (b) The relative change in photocurrent due  to the structures as a function of angle of incidence.}
		
		\end{figure}
	
	To test real-world applicability, we finally simulated the yearly energy harvest of a solar cell with and without triangles and with different tracking strategies using LightTools’ built-in solar simulator that utilizes historical data for direct and diffuse radiation at different locations and times of day and year. The simulated solar cell is $100\mathrm{cm}^2$ square, with three $1.25\mathrm{mm}$ busbars spaced $25\mathrm{mm}$ apart and $75\mathrm{\mu m}$ finger contacts with pitch $2$mm for a metal coverage of 7.5\%. The triangles are $15\mathrm{\mu m}$ above the contacts to account for a hypothetical adhesion layer. The busbar triangles are $2.95$mm wide and $8.7$mm tall. The finger triangles are $245\mathrm{\mu m}$ wide and $730\mathrm{\mu m}$ tall. The overall encapsulation thickness is $9$mm. The cell was simulated in Phoenix, AZ for a year. In the stationary and single-axis tracking simulations the cell is assumed to be inclined 26 degrees to the south. The results are compiled in table 1. We see that enormous increases in yearly energy production could be possible by implementing triangular voids above the contacts and that the enhancement is not dependent on the module’s tracking strategy. In addition, the increase in energy harvest was independent of cell orientation. The increase in energy production being greater than the metallized area and the east/west and two-axis tracking performing differently is a simulation artifact due to modeling a single cell where the edges of the module are important.
	\begin{table}
	\begin{tabular}{|c|c|}
		\hline 
		Tracking Strategy & \% Increase In Annual Energy Relative to Plain Cell \\ 
		\hline 
		None & 7.6 \\ 
		\hline 
		East-West & 8.0 \\ 
		\hline 
		Two Axis & 8.1 \\ 
		\hline 
	\end{tabular} 
	\caption{Simulated annual improvements in annual energy harvest for a solar panel equipped with perfect effectively transparent contacts}
	\end{table}
\section{Experiment}
	We also performed proof of principle experiments demonstrating reduction of reflection losses due to the busbar of a commercially available solar cell. Five inch monocrystalline solar cells were obtained from Amazon with busbar width 1.5mm and a negative mold for the encapsulation was 3D printed using an Object Eden 260VS. The mold is a square basin 2.5mm deep and 25mm across with a triangle running down the middle of width 1.5mm and height 1.5mm. The encapsulation is formed by casting polydimethylsiloxane (PDMS) and letting it cure, and then brushing a thin layer of uncured PDMS onto the interior of the triangle to smooth out roughness due to the finite resolution of the printed mold. The block is then placed on top of a solar cell that has been coated with a thin layer of uncured PDMS to form a good optical contact and then letting the rest of the PDMS cure. Figure 4a is a photograph of the completed structure. A thin white line is all that can be seen of the busbar. This indicates light from the busbar is unable to reach the camera due to scattering by the triangle. Reciprocity then implies that normally incident light is incident on the silicon instead of the busbar. The finger contacts appearing to run through the triangle are reflections of the contacts off the triangle's surface. Figure 4b is a reflection micrograph taken by a microscope with a 5X lens and numerical aperture 0.13. A strongly reflecting finger contact is visible while the busbar on the left is obscured by the triangle. This is further evidence that light is being redirected into the silicon. Light striations in the image along the busbar are visible and are due to the finite layer height resolution of the 3D printer that defined the mold.  Figure 4c is a spatially-resolved laser beam induced current (LBIC) measurement utilizing a 543nm laser. The image is normalized so that the finger contacts are not photoactive and that some silicon is just saturated. The triangle is so effective that the edge of the triangle is only discernable by a jump in the response of the finger contact. The finger contact appears to run through the triangle because light reflecting off the triangle then reflects off the finger contact. Figures 4d,e,f are the analogous images taken of a cell with plain encapsulation. The contrast between the images clearly demonstates the efficacy of the triangles. 
	\begin{figure}[h!]
		\centering
		\def\svgwidth{\columnwidth}
		\begin{framed}
		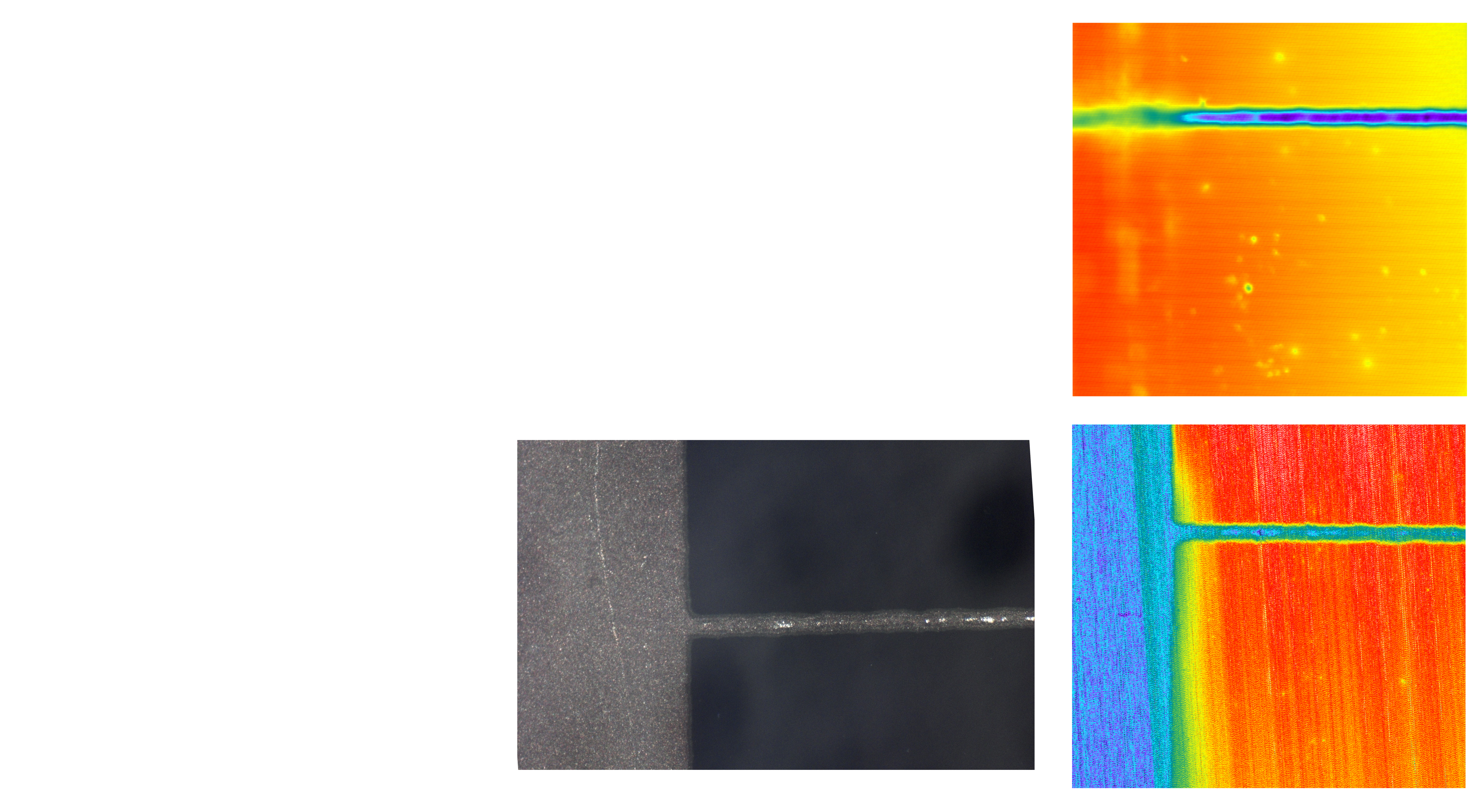
		\end{framed}
		\caption{Experimental demonstration of functioning effectively transparent contacts for a solar cell busbar. (a) Photograph of solar cell busbar with triangular void in PDMS above the busbar. (b) Reflection microscopy image of a busbar with a triangular void above it and a finger contact without a triangle. (c) Spatially resolved laser beam induced current measurement of a busbar with a triangular void above it and a finger contact without a triangle. The dashed line is the approximate location of the busbar. The color indicates the measured photocurrent, scale is arbitrary units. (d,e,f) Images analogous to (a,b,c) except the PDMS layer is planar.}
	\end{figure}
\section{Discussion and Conclusion}
	The presented results suggest triangular voids above the contacts could be a powerful way to reduce reflection losses due to the front contacts of a solar cell. However, many questions must be addressed before this technique is commercially viable. Foremost, high-quality master molds for the encapsulation must be fabricated. The busbars are easily dealt with using traditional machining techniques. The finger contacts, due being about 100 microns wide, and the mold height necessarily being several hundred microns tall, would be difficult to machine, and are too large to be defined using, for example, inductively coupled plasma etching of silicon due to the etch always operating in the infinitely-wide-trench regime, resulting in uncontrollably re-entrant etch profiles.\cite{Henry2010} Anisotropic wet-etching of silicon would excessively limit the triangle aspect ratio. Non-conventional lithographic techniques such as grayscale and inclined exposure may be promising ways forward.\cite{Han2004,Rammohan2011} Another interesting question is how these triangles would affect module reliability. For example, would the gases in the triangle degrade the metallic contacts over time, or possibly ease delamination of the module? 
	
	In conclusion, we have presented a new strategy for eliminating reflection losses due to the front contacts of a solar cell. By placing triangles with an index of refraction less than the surrounding encapsulation above the contacts, total internal reflection can direct the light away from the contact and into the solar cell absorber. Simulations of triangles and other shapes show that the photocurrent can be enhanced over a wide and controllable range of angles of incidence, and simulation of a cell over a year demonstrates that large gains in yearly energy harvest are possible. Proof of principle experiments utilizing readily available components and LBIC demonstrate that this technique does enhance photocurrent and can be extremely simple to implement. 
\section{Acknowledgments}
The authors would like to thank John Lloyd for helpful discussions while setting up the LightTools simulations.

This material is based upon work primarily supported by the Engineering Research Center Program of the National Science Foundation and the Office of Energy Efficiency and Renewable Energy of the Department of Energy under NSF Cooperative Agreement No. EEC‐1041895. Any opinions, findings and conclusions or recommendations expressed in this material are those of the author(s) and do not necessarily reflect those of the National Science Foundation or Department of Energy.
\bibliography{ETC}
\end{document}